\documentclass[preprint,12pt]{elsarticle}

\usepackage{amssymb}
\usepackage{amsmath}

\usepackage{lineno}
\usepackage{subfig}
\usepackage{xcolor}
\usepackage{mathrsfs}

\journal{Nuclear Physics A}

\begin{document}


\begin{frontmatter}



\title{Measurement of the Resolution of the Timepix4 Detector for $100~$keV and $200~$keV Electrons for Transmission Electron Microscopy}


\author[Oxford,QD]{N. Dimova\corref{cor1}}
\cortext[cor1]{Corresponding Author: nina.dimova@physics.ox.ac.uk}
\author[Oxford]{R. Plackett}
\author[Oxford]{D. Weatherill}
\author[Oxford]{D. Wood}
\author[QD]{L. O'Ryan}
\author[QD]{G. Crevatin}
\author[RFI]{J. S. Barnard}
\author[RFI]{M. Gallagher-Jones}
\author[Oxford]{D. Hynds}
\author[QD]{R. Goldsbrough}
\author[Oxford]{I. Shipsey}
\author[Oxford]{D. Bortoletto}
\author[Oxford,RFI]{A. Kirkland}

\affiliation[Oxford]{organization={University of Oxford},
            addressline={Department of Physics, Denys Wilkinson Building}, 
            city={Oxford},
            postcode={OX1~3RH},    
            country={United Kingdom}}

\affiliation[QD]{organization={Quantum Detectors Ltd.},
            addressline={R103, Harwell Campus}, 
            city={Didcot, Oxfordshire},
            postcode={OX11~0QX}, 
            country={England}}

\affiliation[RFI]{organization={The Rosalind Franklin Institute},
            addressline={Harwell Campus}, 
            city={Didcot, Oxfordshire},
            postcode={OX11~0QS},    
            country={England}}

\begin{abstract}
We have evaluated the imaging capabilities of the Timepix4 hybrid silicon pixel detector for $100~$keV and $200~$keV electrons in a Transmission Electron Microscope (TEM). Using the knife-edge method, we have measured the Modulation Transfer Function (MTF) at both energies.  Our results show a decrease in MTF response  at Nyquist (spatial) frequency, dropping from approximately $0.16$ at $100~$keV to $0.0046$ at $200~$keV. However, by using the \emph{temporal} structure of the detected events, including the arrival time and amplitude provided by the Timepix4, we enhanced the spatial discrimination of electron arrival. This approach improved the MTF at Nyquist by factors of $2.12$ for $100~$keV and $3.16$ for $200~$keV. These findings demonstrate that the blurring effects caused by extended electron trajectories within the sensing layer can be partially corrected in the image data.
\end{abstract}




\end{frontmatter}

\newpage 

\section{Introduction}

The Medipix family of readout ASICs~\cite{MEDIPIX} have been designed to enable the use of hybrid silicon pixel detector technology, originally developed for particle physics applications, in other research fields. This technology allows for high-rate, high-resolution radiation imaging and particle tracking.

The Timepix4~\cite{TPX4} is the latest generation of Medipix readout ASICs, developed by the Medipix4 Collaboration. The Medipix2 and Medipix3, detectors with the same pixel pitch and sensor technology as the Timepix4, were quickly established in Electron Microscopy as fast electron diffraction cameras capable of high dynamic range and rapid acquisition speed~\cite{Glsg1}. They were essential in recovering the scattered electron wave-function through ptychography~\cite{ptych} and mapping out crystallographic information in beam sensitive materials~\cite{MOF}, and have been extensively used for imaging in electron microscopes~\cite{E1, E2}. Their development within Particle Physics has allowed these detectors to have both high frame rate and signal-to-noise ratio and so provide sufficient bandwidth for high-flux electron imaging~\cite{FLUX}. Their intrinsic imaging resolution has been demonstrated to be suitable for these applications. Therefore, they serve as a direct comparison for the performance of the Timepix4.

\begin{figure}[!h]
	\centering
	\includegraphics[width=0.8\textwidth]{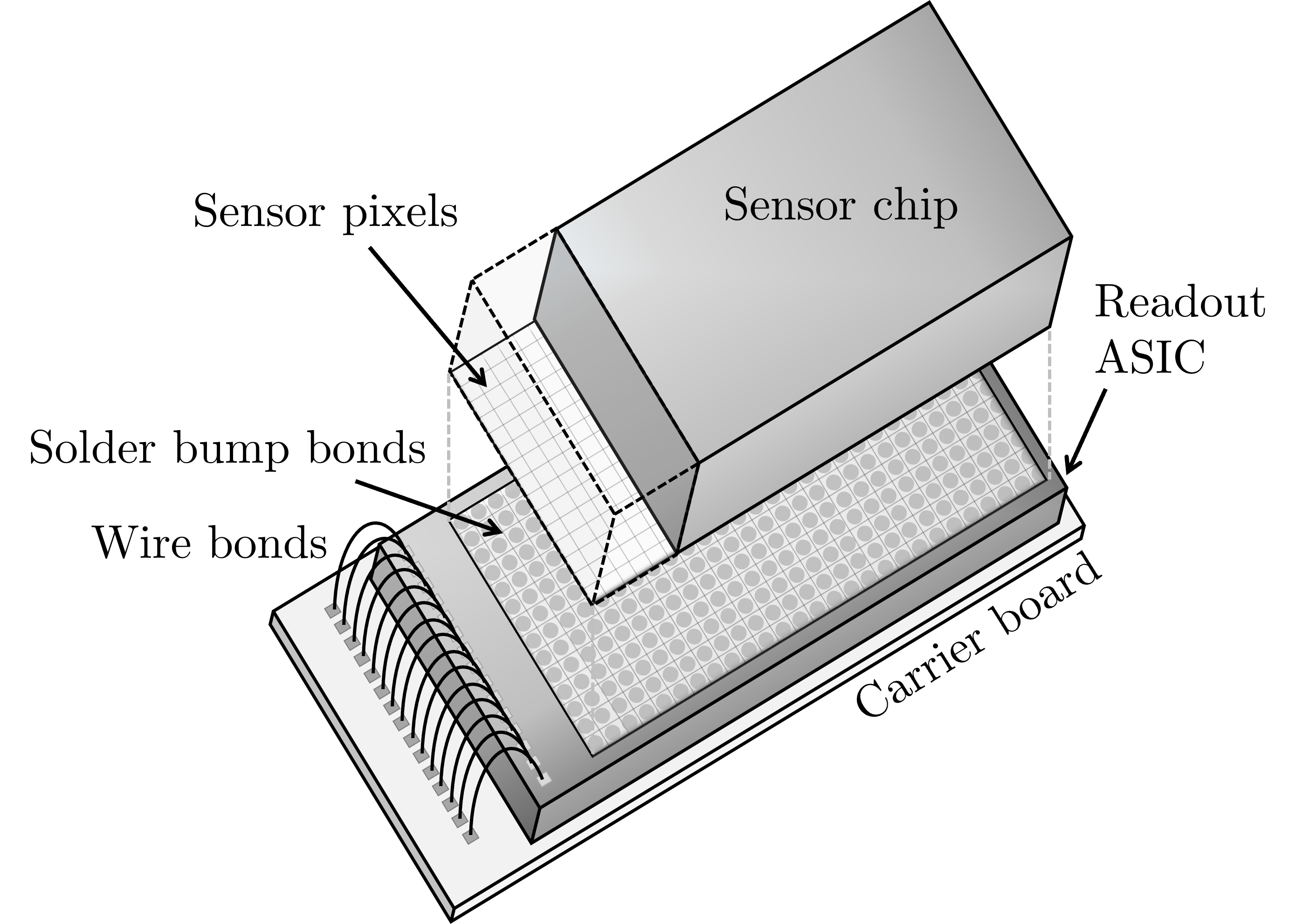}
	\caption{A schematic of a hybrid silicon pixel detector. Not to scale.}
	\label{F1}
\end{figure}

Hybrid silicon pixel detectors consist of two separate silicon chips, a `sensor' containing an array of pixelated reverse-biased diodes made from high-resistivity silicon, and a corresponding grid of amplifiers and digitisers in a `readout ASIC’. The two chips are connected via an array of small solder bump bonds that join the individual sensor diodes to a readout channel (Figure~\ref{F1}).

We utilize a Timepix4 ASIC bonded to a $300~\mu$m thick p-on-n silicon sensor, and read out by a prototype readout system provided by Quantum Detectors Ltd.~\cite{QD}. The Timepix4 is a $24.7~\text{mm} \times 30.0~\text{mm}$ hybrid pixel ASIC produced using $65~$nm CMOS technology. It consists of an array of $448 \times 512$, $55~\mu$m pixels which can be bump-bonded to a pixelated sensor. The analogue front-end of the chip amplifies and converts charge deposited in the sensor by an incident particle into a shaped voltage pulse. The Timepix4 operates in a `data-driven mode', sending out a $64$-bit data packet when a pixel registers a hit above a pre-defined and programmable threshold. This data is read out in two separate streams, one for each half of the pixel matrix.

The data packet generated by the Timepix4 includes the pixel coordinates for the hit, Time of Arrival (ToA) and Time over Threshold (ToT). The ToA records the arrival time of charge at the readout cathode relative to an ultra-fine $5.12~$GHz clock phase. The ToT measures the duration (in $40~$MHz master clock counts) for which the signal remains above the threshold, providing a measure of the charge collected by the pixel. Consequently, the hits are timestamped in $195~$ps time bins, and the ToT is recorded with a resolution of approximately $700~$e$^-$ Full Width Half Maximum ($\sim1.5~$keV FWHM in Si).

The per-hit ToA and ToT data enable further enhancement of spatial resolution. By using the ToA and ToT information, offline data processing can correct for particles that register across multiple pixels; reconstructing these hit clusters allows us to achieve sub-pixel resolution. The high readout rate of the Timepix4 generates a large volume of data compared to previous ASIC generations, necessitating a custom, high-speed, and memory-efficient clustering algorithm.

\begin{figure}[!h]
	\centering
	\includegraphics[width=0.9\textwidth]{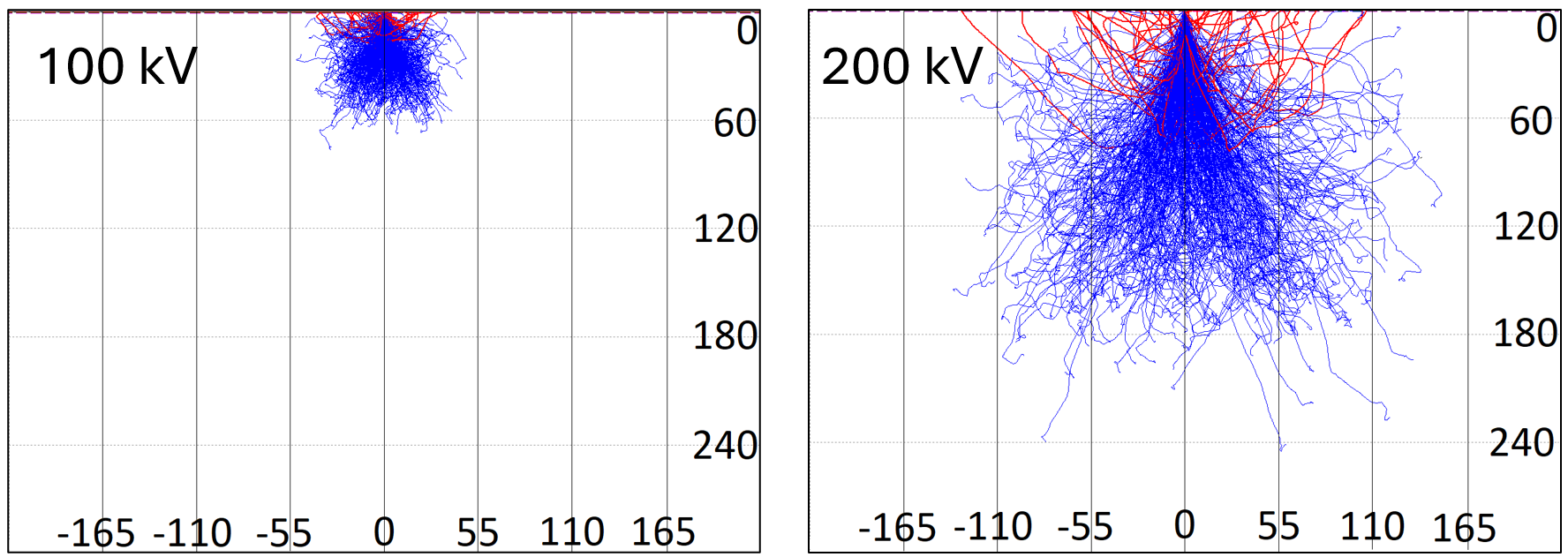}
	\caption{Monte-Carlo simulations~\cite{MC} of one thousand $100~$keV (left) and $200~$keV (right) electrons incident on the Timepix4 sensor (surface contact layers have been included). The depth and lateral dimensions are in micrometres ($\mu$m) with black vertical lines indicating the 55$~\mu$m pixel spacing; blue tracks mark fully absorbed electrons, red tracks mark electrons that backscatter out of the sensor,  resulting in reduced energy deposition.}
	\label{F9}        
\end{figure}

Improving the base resolution of the detector is particularly important in transmission electron microscope (TEM) applications. Unlike X-ray photons, electrons generate a large ionisation region inside silicon, often producing signals across multiple pixels, particularly in thicker sensors where electron scattering is significant. This effect makes it challenging to accurately pinpoint their entry location on the sensor, as shown in Figure~\ref{F9}. The relatively large pixel size further complicates the precise determination of the electron entry point.

\section{MTF and the slanted knife-edge method}
\label{S2}

The Optical Transfer Function (OTF) characterises an imaging system by describing its response to a sine wave input, as a function of its spatial frequency and orientation. It provides a full description of the system's response, including phase effects, and is defined as the Fourier transform of a Point Spread Function (PSF). Essentially, the OTF describes how well the system can image a point source.

This work focuses on the Modulation Transfer Function (MTF), which represents the magnitude of the OTF. Due to the sensor’s square pixels, the system exhibits high symmetry and does not capture any phase information, making the MTF particularly relevant. The MTF can also be defined in terms of the contrast of an object, and the contrast of its image (Eq.~\ref{E1}): 
\begin{align}
	C &= \frac{I_{max} - I_{min}}{I_{max} + I_{min}} \nonumber \\[6pt]
	MTF(\omega) &= \frac{C_{image}(\omega)}{C_{object}(\omega)} \label{E1}\\
    \nonumber
\end{align}
The MTF is a standardised metric~\cite{ISO} for comparing optical and other imaging systems, detailing their ability to convey contrast as a function of spatial frequency. Typically, the MTF is shown as a function of frequency, normalized to the system’s Nyquist frequency defined as half the sampling rate (0.5 cycles per pixel for a pixelated detector). This measure, the MTF at Nyquist frequency, is a widely used parameter for characterizing imaging detectors ~\cite{MTF1, MTF2}.

\begin{figure*}[!h]
	\centering
    \scalebox{1.3}{
    $\text{ESF} \xrightarrow[\text{}]{\text{differential}} \text{LSF} \xrightarrow[\text{}]{\text{Fourier transform}} \text{OTF} \xrightarrow[\text{}]{\text{magnitude}} \text{MTF}$} \\
	\subfloat{\includegraphics[width=0.33\textwidth]{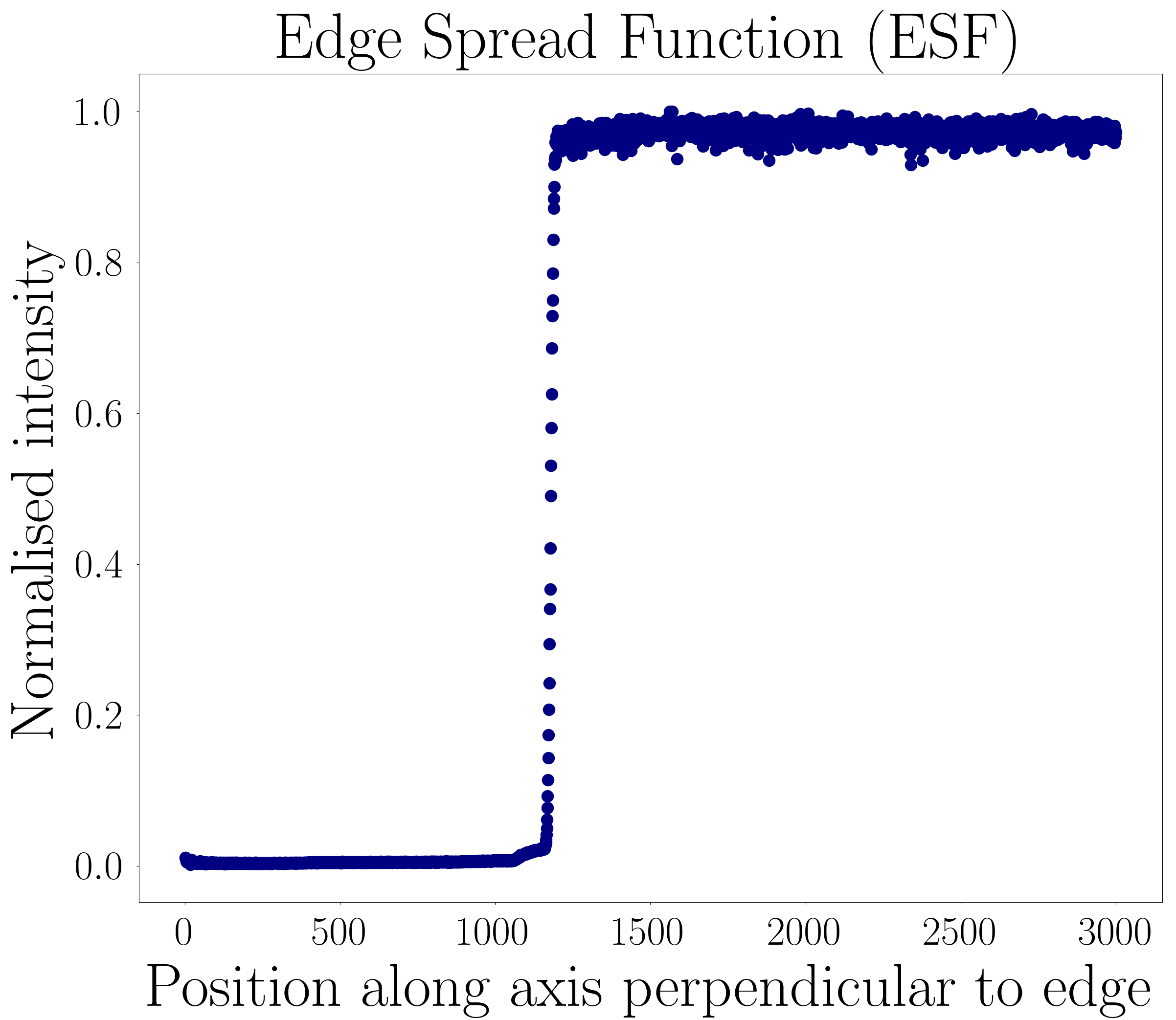}}
    \subfloat{\includegraphics[width=0.337\textwidth]{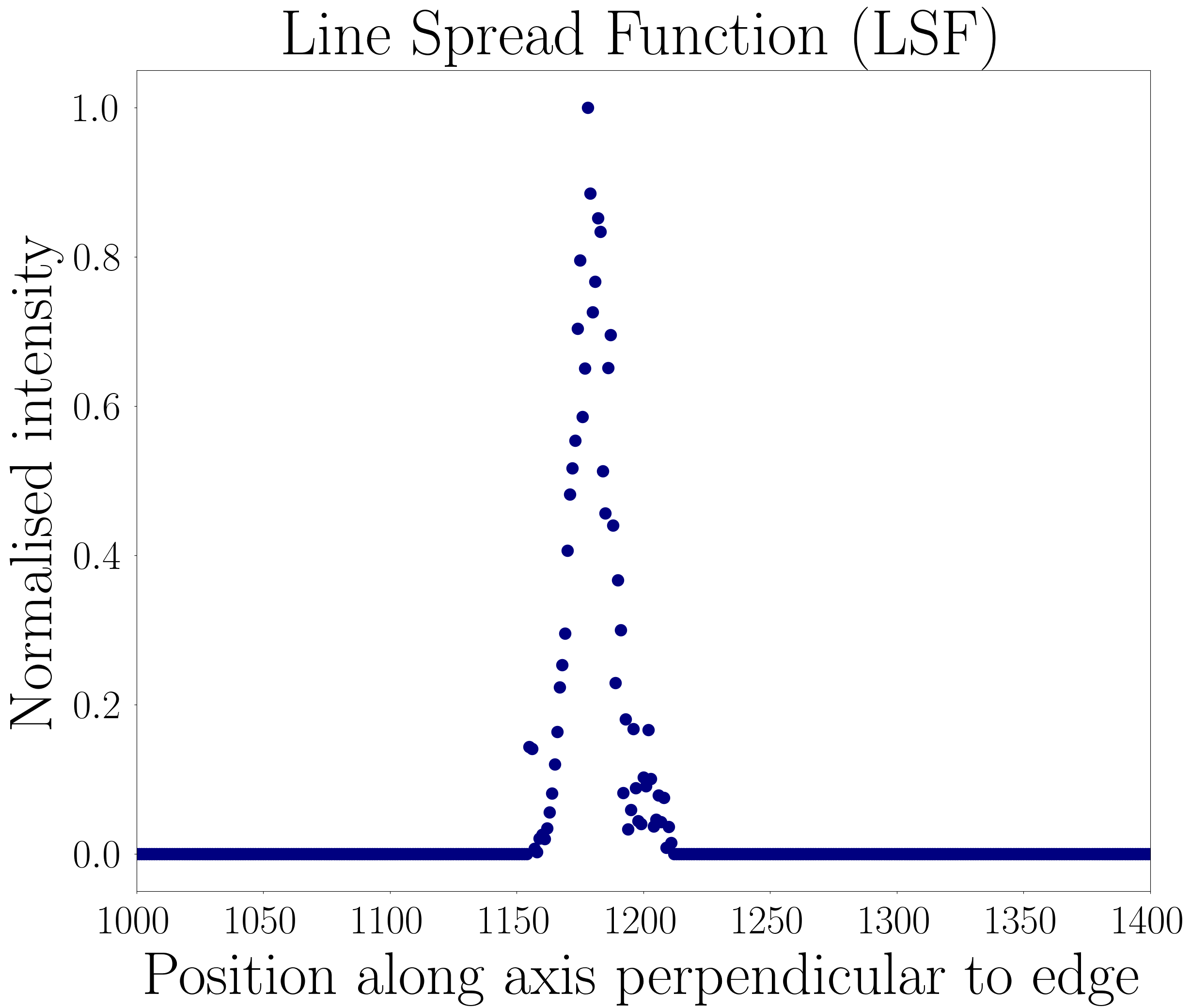}}
    \subfloat{\includegraphics[width=0.333\textwidth]{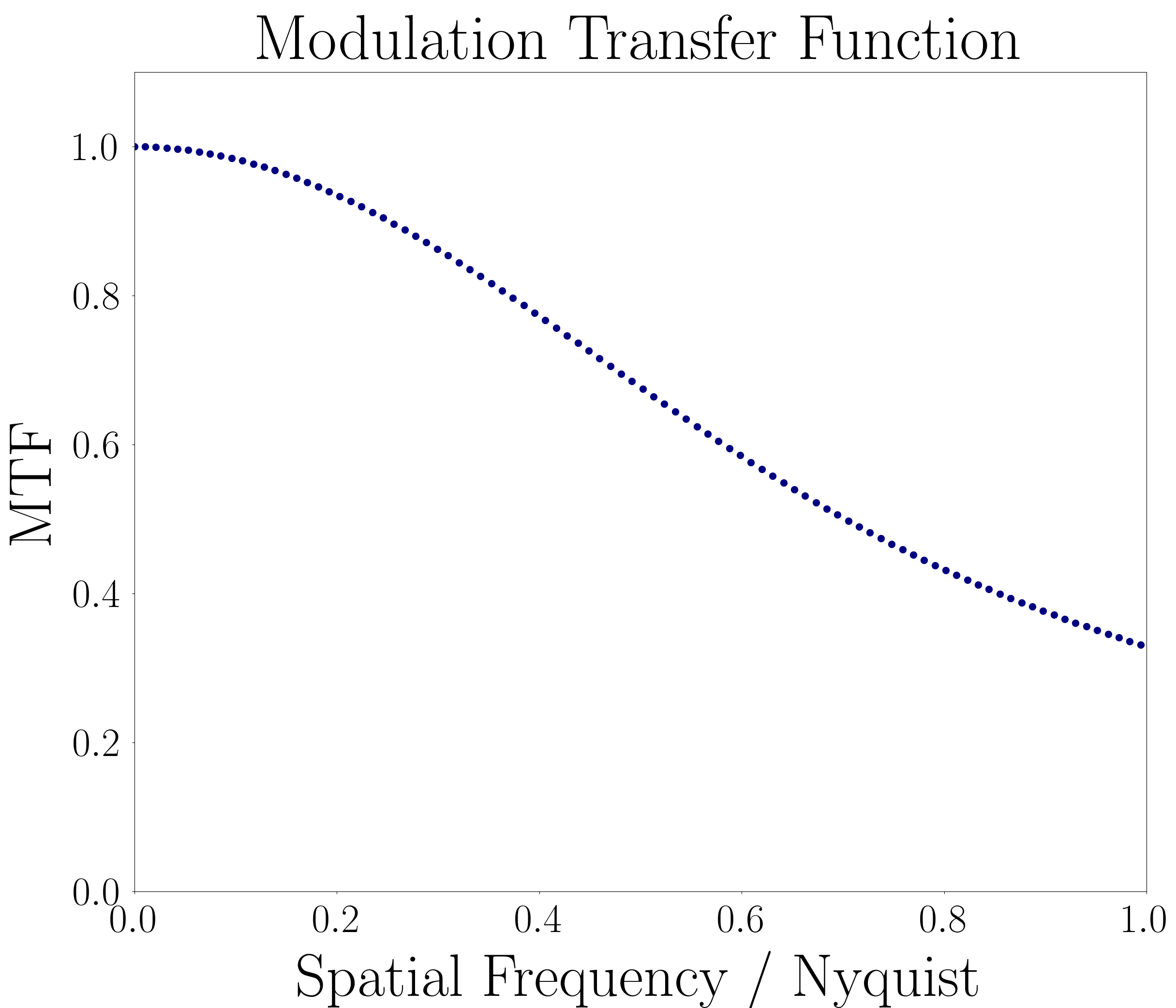}}
	\caption{An illustration of the MTF calculation method. The figures show example ESF, LSF, and MTF obtained from real data. Before differentiating, the high-intensity part of the ESF has been averaged to reduce noise contributions to the LSF.}
	\label{F2}
\end{figure*}

In optical systems, the 2D OTF, which describes the detector's response in all directions, is derived as the Fourier transform of the PSF. However, in pixelated detectors, most of the non-zero portion of the PSF is contained within a single pixel. To address this, a 1D section of the OTF can be obtained by performing a Fourier transform of the Line Spread Function (LSF). The LSF represents the detector's response to a thin slit projected onto the sensor. However, the data collected through a very narrow slit is limited by the sharpness of the slit. To overcome this limitation, the integral of the LSF is used, known as the Edge Spread Function (ESF). The ESF reflects the detector's response to a knife-edge projected onto the sensor, and the differential of the ESF provides the LSF (Figure~\ref{F2}).

For comparison, the MTF of a perfect pixel with pixel pitch $d$ is given as the Fourier transform of the pixel intensity response, a top-hat function, which is a sinc function of the spatial frequency: $\mathscr{F}\left(\sqcap(x/d)\right) \propto \text{sinc}(\omega d).$

When the input signal contains frequencies above the Nyquist frequency of the detector aliasing can occur. Frequencies beyond Nyquist are ``folded back" onto lower frequencies, making it impossible to distinguish between aliased and non-aliased signals in general cases.

\begin{figure}[!h]
	\centering
	\includegraphics[width=0.8\textwidth]{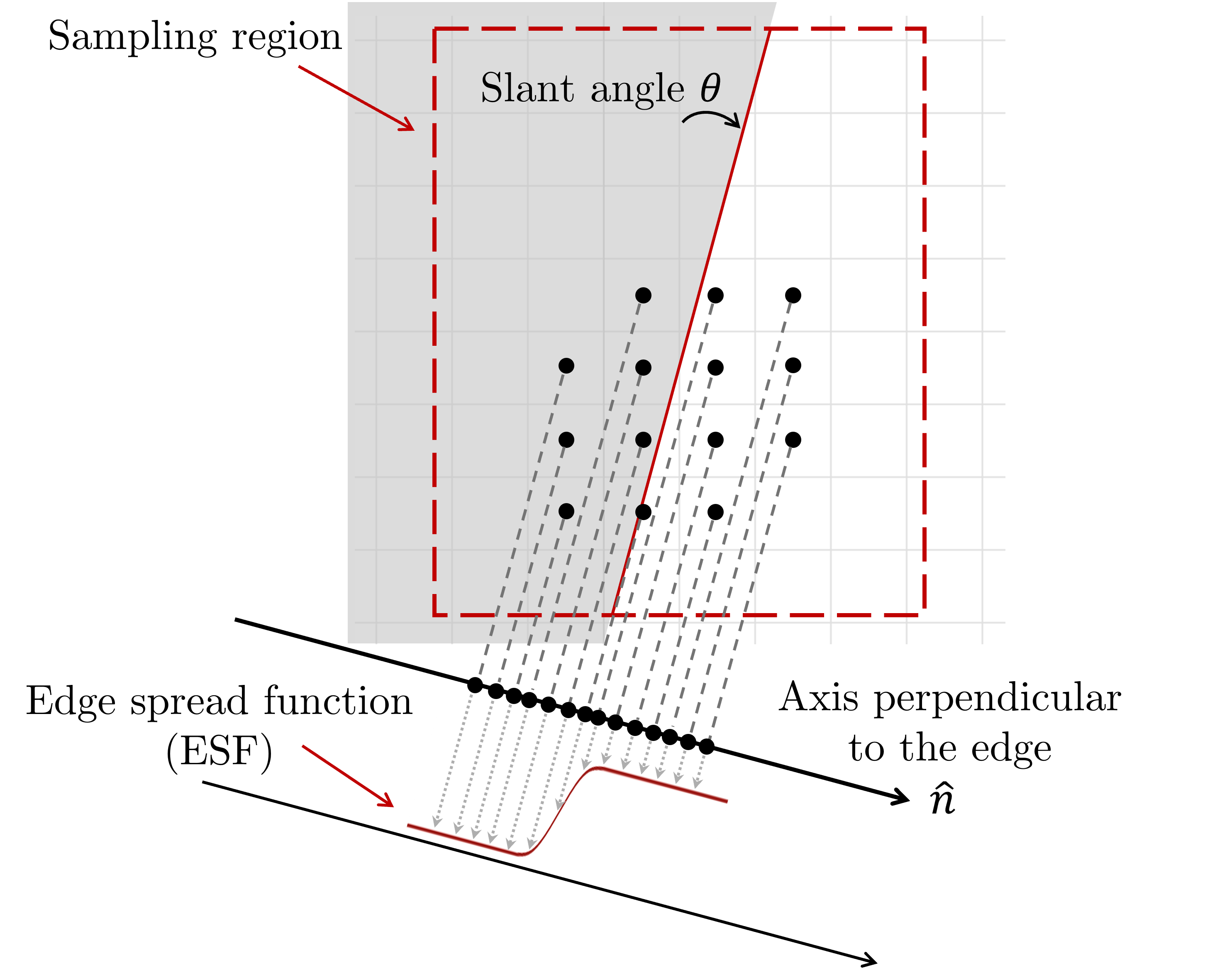}
	\caption{An illustration of the oversampling procedure achieved by slanting the knife edge with respect to the pixel grid.}
	\label{F3}
\end{figure}

One way to prevent aliasing is through oversampling. By sampling a signal multiple times at different points in its cycle and combining those samples, a higher effective sampling rate can be achieved.  In pixel detectors, each pixel row  can act as a distinct sample of an edge. If the edge is perpendicular to the pixel grid, each row will sample it at the same point. However, if the edge is slanted with respect to the pixel grid, each row gives a sample at a different point in the edge cycle (Figure~\ref{F3}). If the slant angle is small, the resulting MTF can be approximated as the horizontal component of a separable 2D MTF, as illustrated in Figure~\ref{F3}. This ``pseudo-MTF" represents the MTF averaged over all positional phases of the incoming radiation, unlike the full 2D MTF that can be directly obtained in optical systems.

To achieve oversampling in practice, an axis perpendicular to the edge is used, and the pixel coordinates are projected along the edge onto this perpendicular axis. This creates a 1D sample with spacing finer than one pixel. The sample is then binned and the chosen binning, combined with the slant angle of the edge, determines the oversampling according to Eq.~\ref{E2}. Although a low slant angle reduces achievable oversampling, it ensures that the 1D projection of the MTF is measured along a horizontal axis parallel to the pixel grid. A larger angle would increase oversampling but introduce contributions from the MTF in the vertical direction.

\begin{equation}
	\text{Oversampling} = \frac{n_{bins}}{\cos\theta}
	\label{E2}
\end{equation}

Modifying the standard ISO 12233:2023 method for MTF calculation~\cite{ISO}, we select $n_{bins} = 8$ resulting in 1D sample bins that are 1/8th of a pixel in size, based on the analyzable area within the pixel grid. We work with angles $\theta \in [3^{\circ}, 10^{\circ}].$

\section{Experimental Methods}

\begin{figure}[!h]
	\centering
	\subfloat[]{\includegraphics[width=0.625\textwidth]{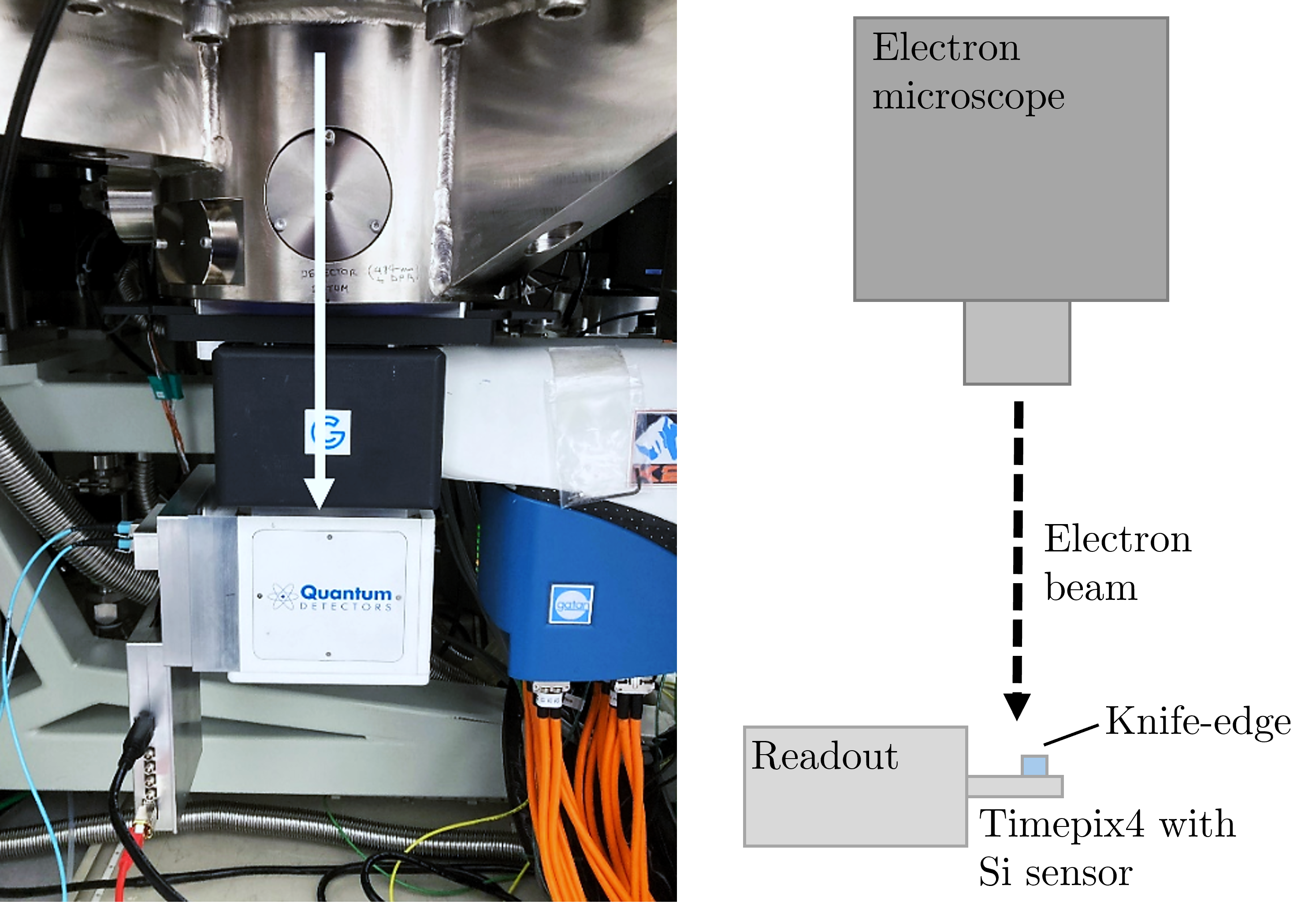}}
	\vfill
	\subfloat[]{\includegraphics[width=0.24\textwidth]{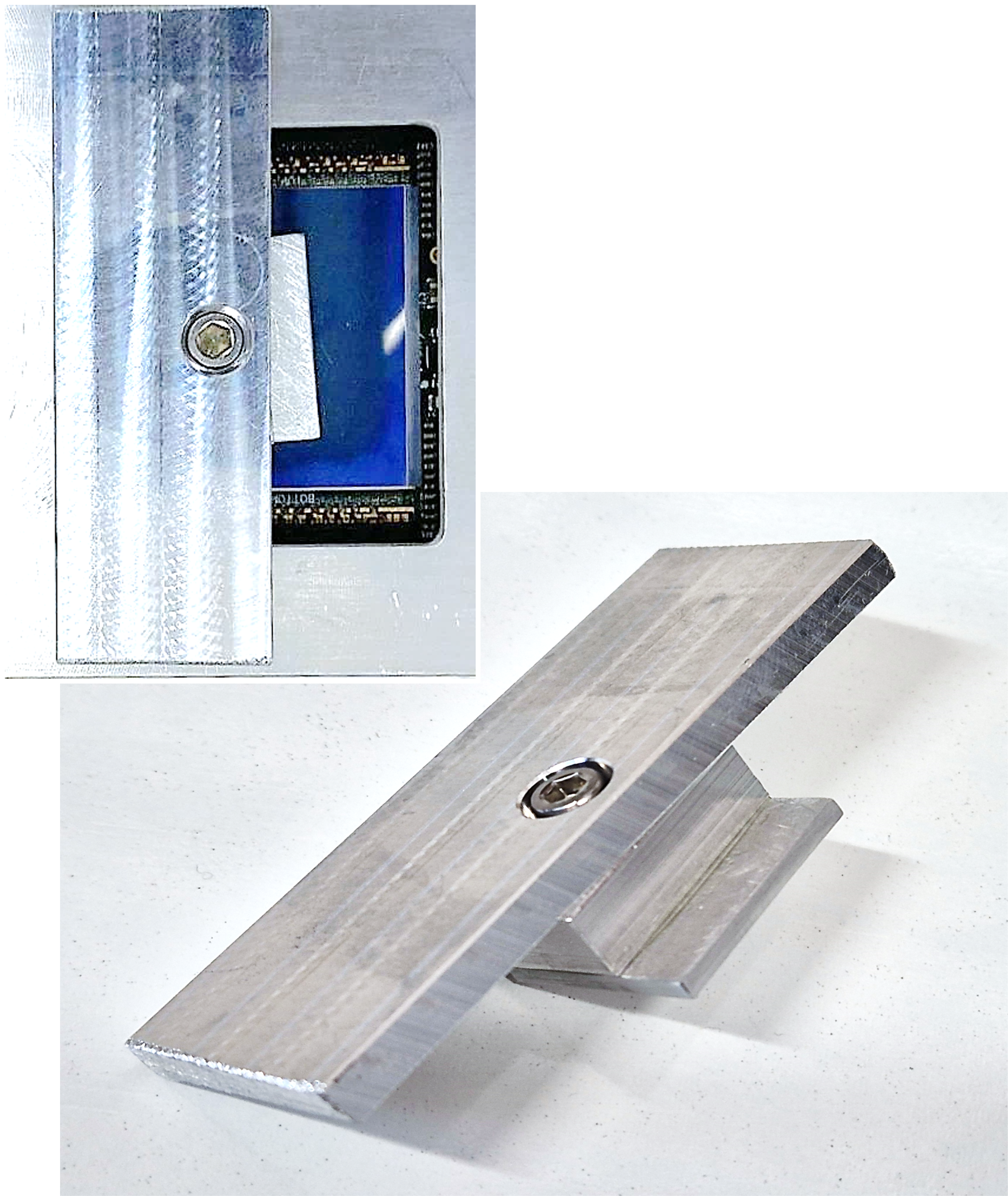}
    \includegraphics[width=0.75\textwidth]{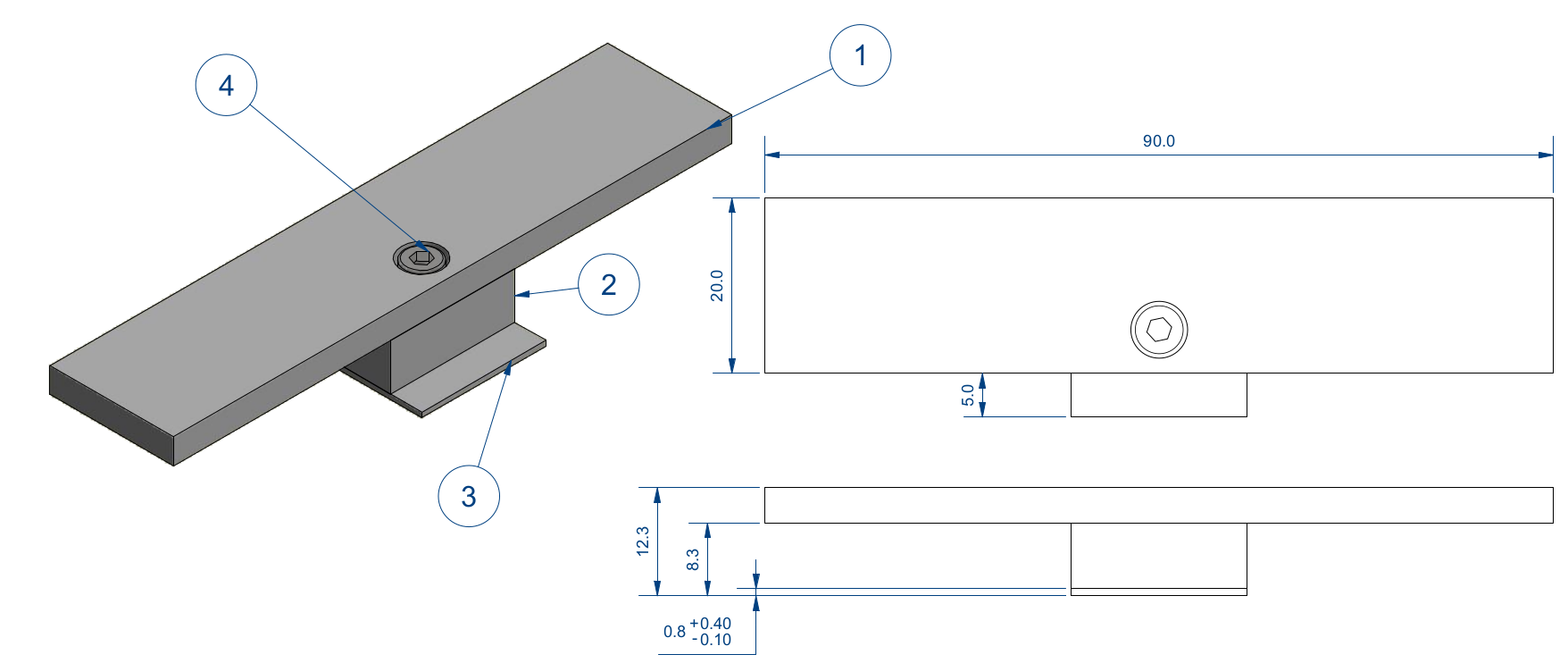}}
	\caption{(a) Data collection with the RFI electron microscope. (b) A custom aluminum knife-edge was fabricated for electron data collection. It consists of a `wing' (1) and a `step' (2, 3) attached to it via a screw (4). First, the step is rotated and fixed roughly at the desired angle with respect to the long side of the wing. Then, the wing is aligned roughly parallel to the pixel matrix and attached to the detector casing. Test data is taken to measure the angle of the edge slope on the sensor and if needed, the step is adjusted again, while the wing remains attached to the detector casing.} The knife-edge is shown in position above the sensor (top-left).
	\label{F4}
\end{figure}

All data were collected using a JEOL Z300FSC cryoARM 300 Transmission Electron Microscope (TEM) with a nominal beam energy of $300~$keV, located at the Rosalind Franklin Institute (see Figure~\ref{F4}). The TEM was operated at $100~$keV and $200~$keV to accommodate the sensor layer’s thickness, which is insufficient for the projected range of a  $300~$keV electron beam (approximately $400~\mu$m). Operating at lower energies also mitigates the risk of radiation damage from the large number of $300~$keV electrons that would otherwise terminate near or penetrate the CMOS layers of the Timepix4 ASIC, and lead to total ionising dose effects.

The Timepix4 sensor assembly was mounted on the TEM using a housing developed by Quantum Detectors Ltd, based on the mechanics of their commercial Medipix3 system (the Merlin camera). The $300~\mu$m planar sensor was controlled by a Zynq UltraScale+ FPGA for slow control links and read out using a Virtex UltraScale+ FPGA via $2.56~$Gbps FireFly links. Only two of the sixteen optic fibre data buses were utilized, limiting the incident current on the sensor to approximately $6~$pA. With all the links active and full connection bandwidth there is scope to raise the particle rate by at least a factor of 32. All current calibrations were made with a Keithley model $6485$ pico-ammeter via BNC connectors to a Faraday cup within the TEM. The sensor was operated in reverse-bias mode at $100~$V (for hole collection), and the pixel response threshold was set to approximately the charge generated by a $10~$keV photon (around $2600$ electron-hole pairs in the sensor).

Two categories of data were recorded with the Timepix4 detector. First, flat-field images were collected by illuminating the sensor with a uniform electron beam such that each pixel received approximately $>10$k hits~\cite{FF}. This ensured that any pixel-to-pixel variations observed were above statistical (Poisson) noise. Second, a slanted, opaque edge was placed just above the sensor to create a knife edge shadow, and images were collected at a rate exceeding 3,000 hits per pixel. This setup used a precision-manufactured aluminum piece positioned 0.8 mm above the sensor (see Figure~\ref{F4}). Fresnel fringe diffraction effects were substantially smaller than the pixel size. The aluminum bulk reduces Bremsstrahlung photon generation~\cite{Brem}. Secondary fluorescence by the aluminum K-X-rays were considered negligible, as they remained well below the Timepix4 energy threshold. Finally, the aluminum edge could be rotated to any desired angle for MTF measurements. In the following, we use an angle $\theta=4^{\circ}$. Additional data were collected and analyzed at $\theta=10^{\circ}$, a large enough angle to introduce contributions from the MTF component perpendicular to the measured direction, as discussed in Section~\ref{S2}. This resulted in a lowered MTF as the sampling region expanded, due to increasing contributions.

\section{Analysis of Results}

To produce an MTF curve for a given dataset, the ESF is first obtained. There are several approaches to obtaining an LSF curve from the ESF. One approach previously employed for MTF measurements~\cite{ERF1, ERF2} is to fit the ESF with an error function
\begin{equation}
    A~\text{erf}\left(\frac{x-\mu}{\sigma}\right) + B,
\end{equation}
where $\mu$ and $\sigma$ are the mean position and width of the function, and $A$ and $B$ are fit parameters. This approach minimizes noise contributions to the measurement. The fitted curve is then differentiated to obtain a Point Spread Function (PSF), which is subsequently Fourier transformed to yield the MTF, with adjustments made for the oversampling rate.

An alternative method to derive the MTF involves averaging the flat regions of the ESF to further reduce noise. The corrected ESF is then directly differentiated—without fitting—to obtain the PSF and, consequently, the MTF.

For $100~$keV electrons, the MTF values at Nyquist obtained with and without fitting the ESF are consistent, which justifies the choice of an error function as a fit. However, due to increased cluster size, $200~$keV electrons create noisy datasets and obtaining the MTF curve without a fit becomes impossible. Therefore, for consistency, the ESF was fitted for all datasets and the LSF and MTF were calculated from the fits. 

\subsection{Particle-counting mode}

\begin{figure}[!h]
	\centering
	\subfloat[]{\includegraphics[width=0.56\textwidth]{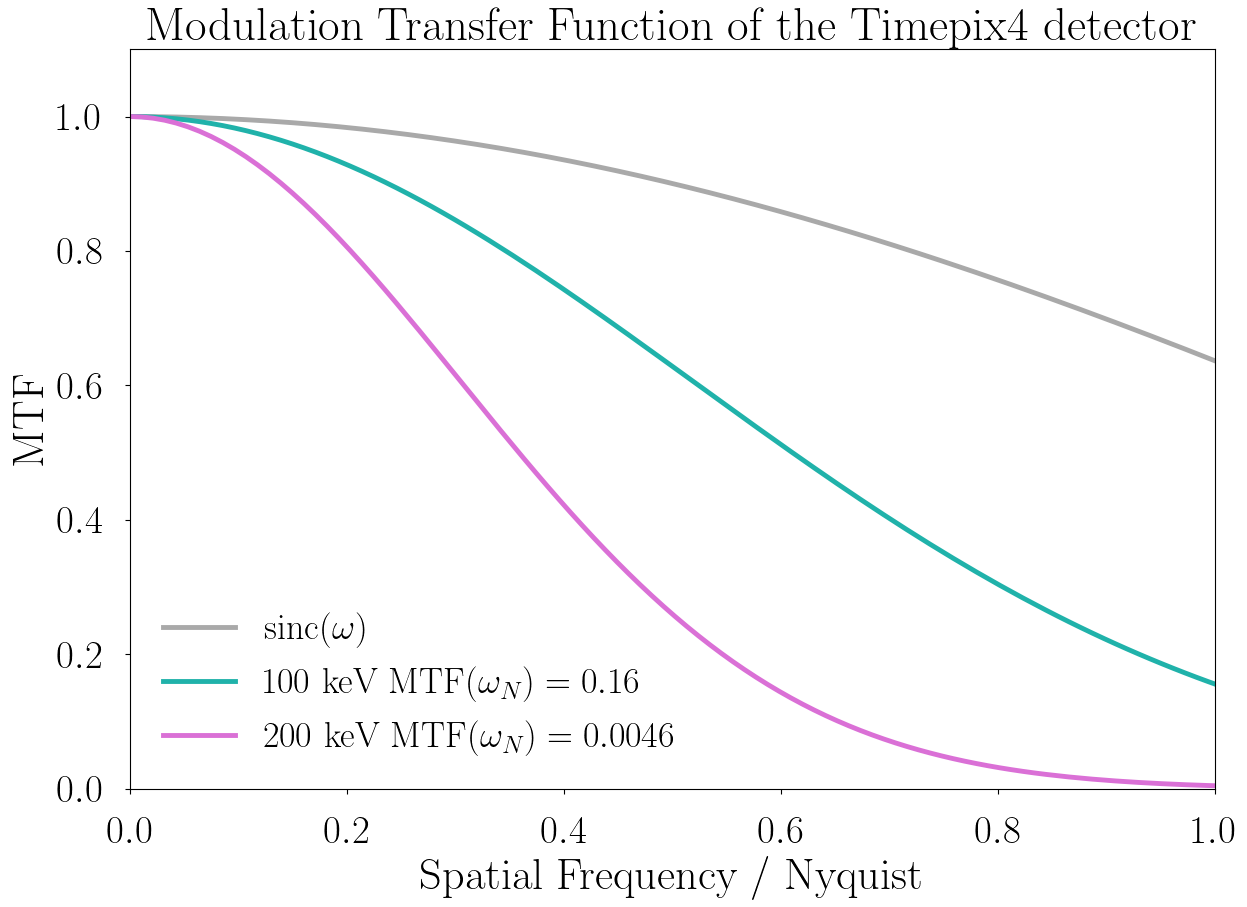}}
    \hspace{0.25em}
    \subfloat[]{\includegraphics[width=0.42\textwidth]{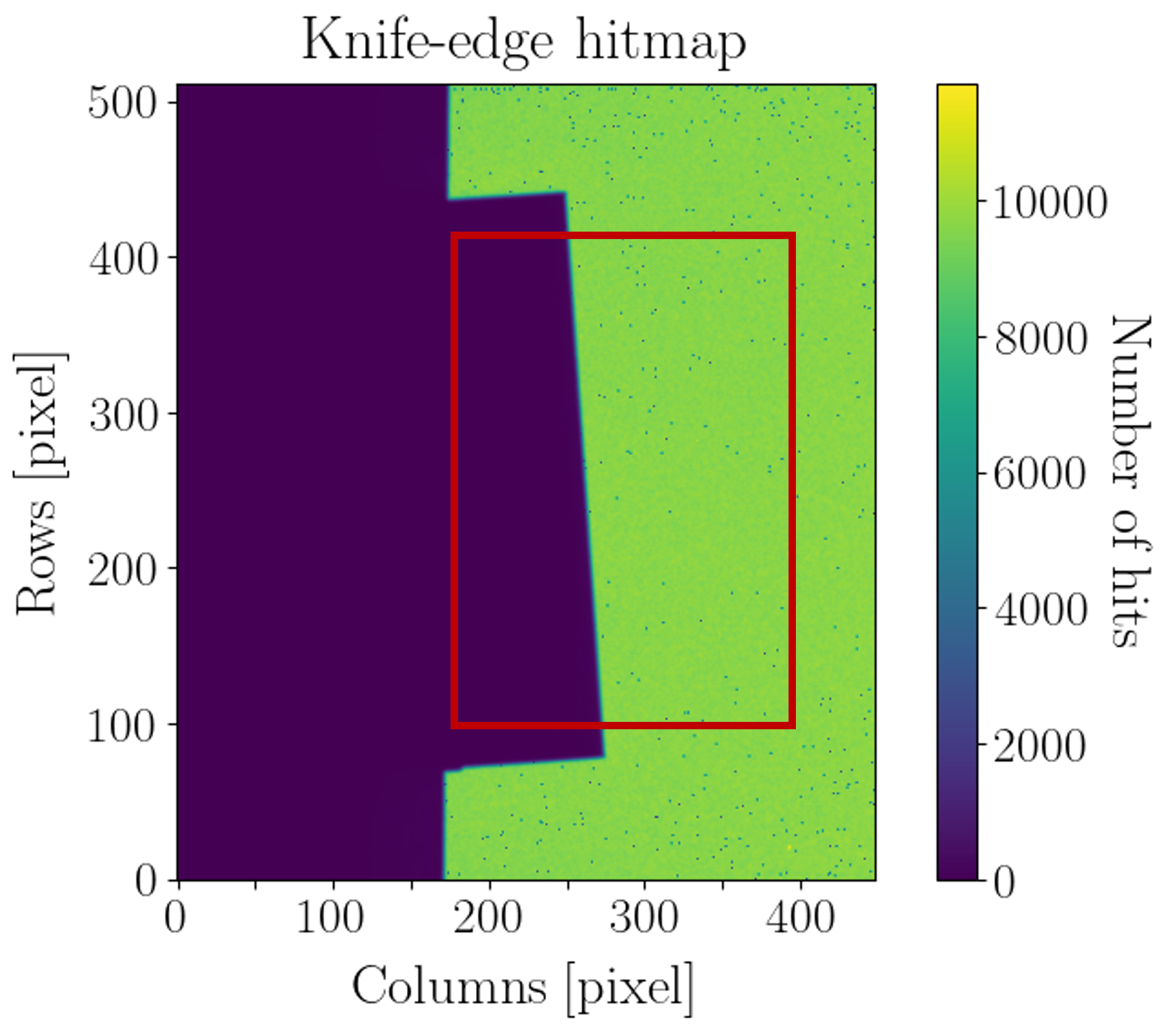}}
	\caption{(a) MTF curves for $100~$keV and $200~$keV electrons, obtained with slant-edge angle $\theta=4^{\circ}$. (b) The sampling region selected for MTF calculation. A large part of the bright area is used to facilitate averaging of the ESF.}
	\label{F5}
\end{figure}

The MTF curves for the Timepix4 were initially computed as if it were operating in a pseudo-particle-counting mode. Only the spatial coordinates of each hit provided by the ASIC were utilized, and the hits within each pixel were summed to generate an image directly comparable to that produced by a particle-counting detector such as Medipix3 (Figure~\ref{F5}). A standard flat-field correction was applied to the particle counting images. The sampling region on the pixel grid selected for the MTF calculation was varied to contain between $200$ and $350$ pixel rows. The values of the MTF at Nyquist were taken as the averages of the values obtained with these sampling regions, effectively correcting for noise contributions to the ESF.

The MTF curves obtained by this method were roughly consistent with transmission electron microscopy characterizations~\cite{TEM1, TEM2} performed previously on Medipix2 and Medipix3 detectors, which were operated with the same sensor type and pixel pitch. More recently, Paton et al.~\cite{TEM3} have measured the MTF values at Nyquist frequency for a $300~\mu$m p-on-n silicon sensor bonded to a Medipix3 ASIC, with a threshold set at $12.4~$keV. They report MTF$(\omega_N)=0.14$ for $120~$keV electrons, and MTF$(\omega_N)=0.01$ for $200~$keV electrons.

\subsection{Clustering}

In addition to spatial hit coordinates, the Timepix4 provides timing and energy information for each recorded pixel hit, specifically the Time of Arrival (ToA) and Time over Threshold (ToT). Hit clustering can be performed using the ToA information to group hits and isolate individual electron events. A ToT-based center of gravity (`centroid’) calculation is then employed to estimate the electron hit position with sub-pixel resolution.

Initially, the recorded hits are sorted by ToA and divided into $100~$ns intervals. This interval was chosen because the drift time of electron-hole pairs across the silicon sensor is on the order of tens of nanoseconds. Within each time interval, hits are clustered spatially; they are considered part of the same event if their coordinates fall within a $7\times7$-pixel window, ensuring that all pixels in a typical electron track are included in the cluster. This clustering algorithm does not filter events based on the cluster shape and so takes all recorded data into account.

Once an event cluster of pixels is established, the coordinates and total ToT of its ToT-weighted centroid are calculated, assigning the entire event to a single sub-pixel coordinate. ToT information is utilized for centroiding as it ensures consistent grouping of pixels, even in cases such as the current work where corrections for the amplifier rise-time effect (`timewalk’)~\cite{TW} have not been applied to the detector. In instances of one-by-$n$-pixel clusters, where $n$ is any integer, there is at least one coordinate for which no additional spatial information is available to estimate the centroid position. In such cases, this coordinate is assigned according to a random uniform distribution within the pixel.

\begin{figure}[!h]
	\centering
	\subfloat[]{\includegraphics[width=0.485\textwidth]{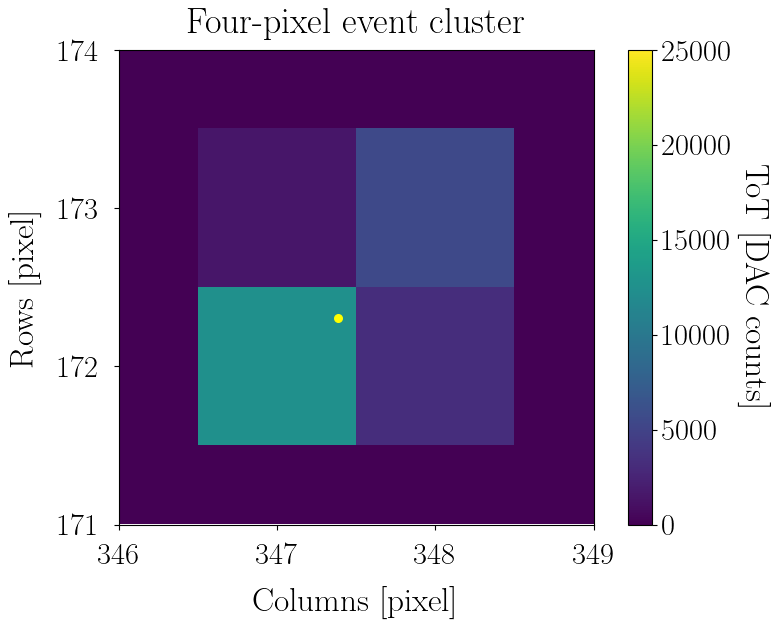}}
    \hspace{0.3em}
    \subfloat[]{\includegraphics[width=0.495\textwidth]{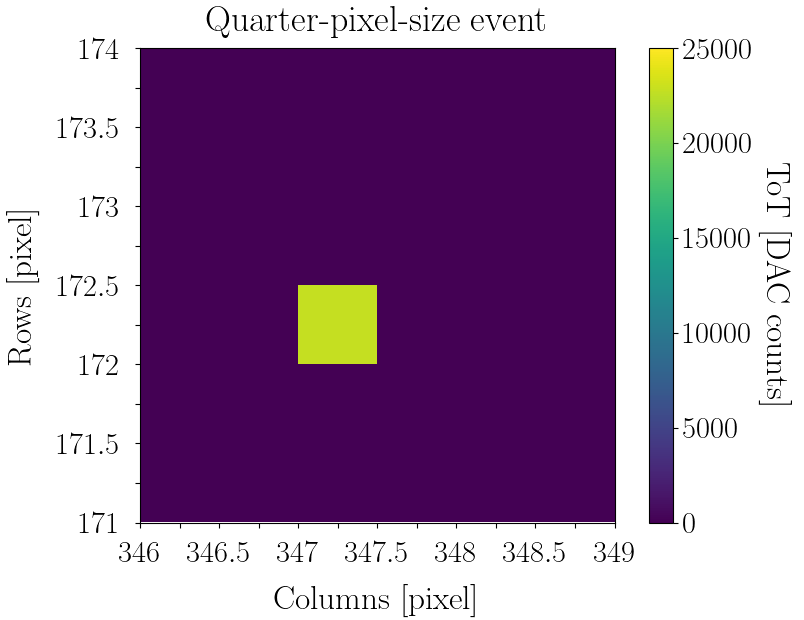}}
	\caption{A visual guide to how the clustering algorithm modifies an electron event cluster (a) into a single hit assigned to a quarter of a pixel (b). The yellow dot is the ToT-weighted centroid of the cluster of pixel hits.}
	\label{F6}
\end{figure}

The pixel is subdivided into a $2\times2$ array of virtual pixels and the hits are assigned based on the calculated cluster centroid locations (Figure~\ref{F6}). This allows for a direct comparison of the MTF calculations between the pseudo-particle-counting mode data presented earlier and the clustered data.

\begin{figure}[!h]
    \centering
	\subfloat[]{\includegraphics[width=0.7\textwidth]{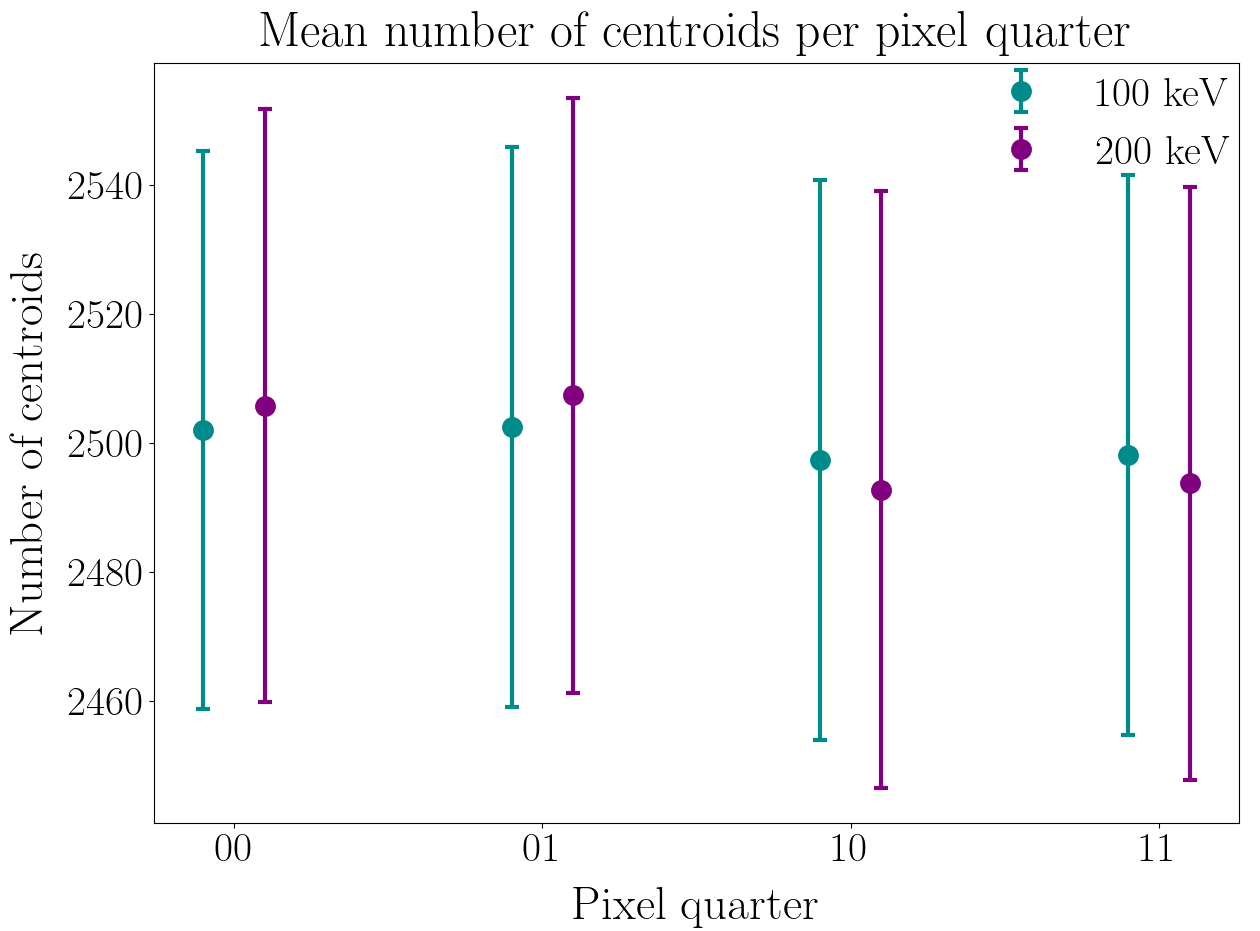}}
    \vfill
    \subfloat[]{\includegraphics[width=0.45\textwidth]{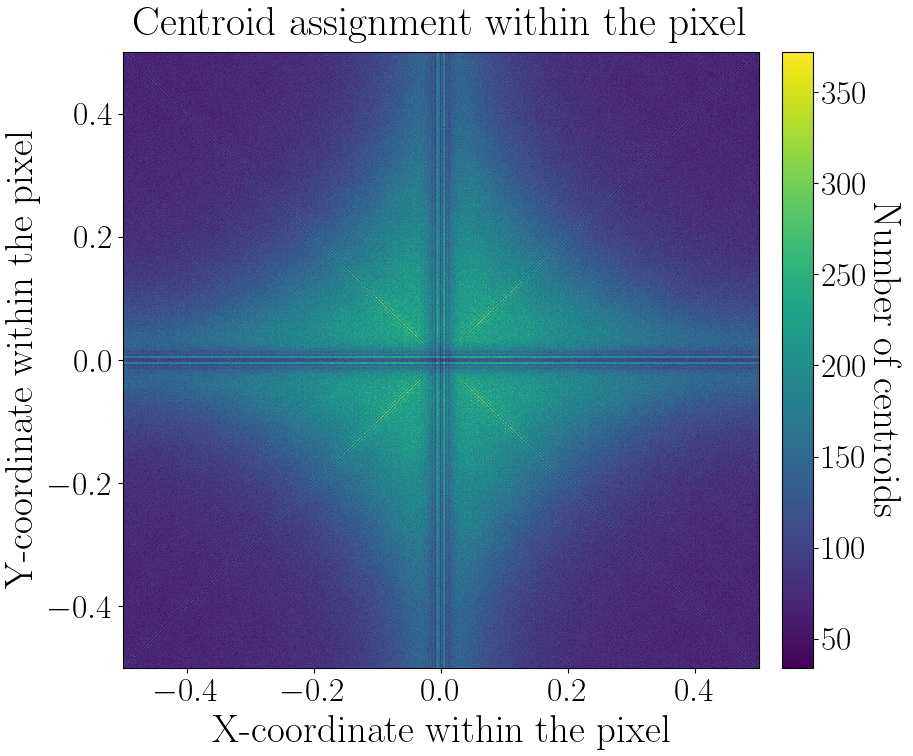}}
    \hfill
    \subfloat[]{\includegraphics[width=0.45\textwidth]{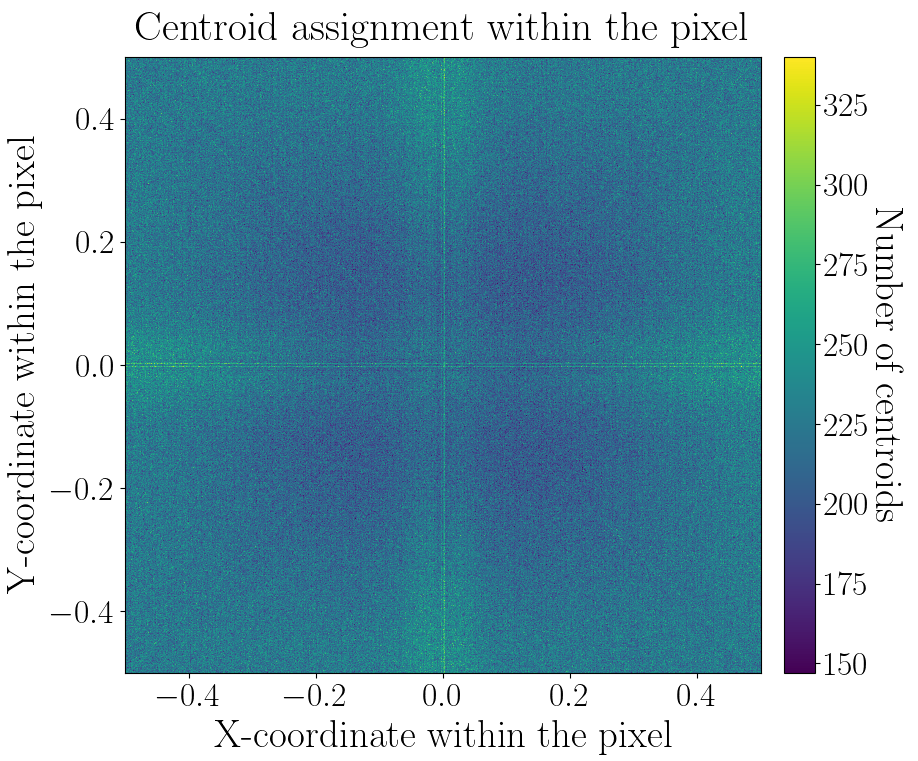}}
	\caption{(a) Mean number of clusters assigned to each of the four virtual pixels within a physical pixel. The data were split into groups of $10$k clustered events to obtain a mean and standard deviation for the whole dataset. The even distribution per quarter demonstrates that the clustering algorithm does not bias the centroid position, up to $2\times2$ splitting of the pixel. The 2D histograms of centroid positions within a pixel for (b) $100~$keV and (c) $200~$keV electrons justify the choice of $2\times2$ splitting. These two histograms include all clusters larger than two pixels. The structure is understood due to the shapes of the observed clusters.}
	\label{F12}
\end{figure}

It is important to examine the distribution of cluster centroids within a pixel divided this manner, in order to ensure that any improvement in resolution achieved through clustering cannot be attributed to potential artificial minimisation of the sensitive area of the pixel, or to biases in the centroid position. Figure~\ref{F12}a illustrates the mean number of centroids assigned to each quarter of the pixel, i.e. to each virtual pixel. The even distribution among the quarters demonstrates that, up to $2\times2$ splitting, the observed improvement in MTF following clustering and re-binning is solely due to an improved estimation of the electron entry point.

Figures~\ref{F12}b and ~\ref{F12}c show 2D histograms of cluster centroid positions within a pixel, for clusters larger than two pixels. The non-uniform distribution is well understood in the context of the possible cluster shapes we observe. For example, 3-pixel clusters in which the pixels form an `L' shape, lead to the centroid being partially offset around the middle of the pixel. These `L'-shaped clusters account for $99.2\%$ of all 3-pixel clusters in the $100~$keV data. In the $200~$keV data, 4-pixel clusters in a $2\times2$ configuration account for $40.1\%$ of all 4-pixel clusters and lead to the offset of the centroid towards the edge of the pixel. The diagonal structure in the $100~$keV data is mostly due to disconnected 3-pixel clusters and symmetric `L'-shaped clusters where the signal in some pixels was not above threshold, which offset the centroid equally along both axes.

These histograms demonstrate why further division of the pixel into smaller virtual pixels is not possible for this electron data. If the pixel were split further, the central virtual pixels would be biased, receiving a disproportionately higher number of centroids compared to those at the pixel's periphery, a phenomenon attributed to the predominant shapes of the clusters observed. For the following resolution analysis a 2 by 2 subdivision of the pixel is used.

The high data output rate of the Timepix4 presents challenges in implementing existing clustering algorithms. Nevertheless, leveraging the MTF datasets already collected, the current clustering method can be optimized, allowing for a comparative analysis of different approaches.

\subsection{Resolution improvement}

\begin{figure}[!h]
    \centering
    \subfloat[]{\includegraphics[width=0.5\textwidth]{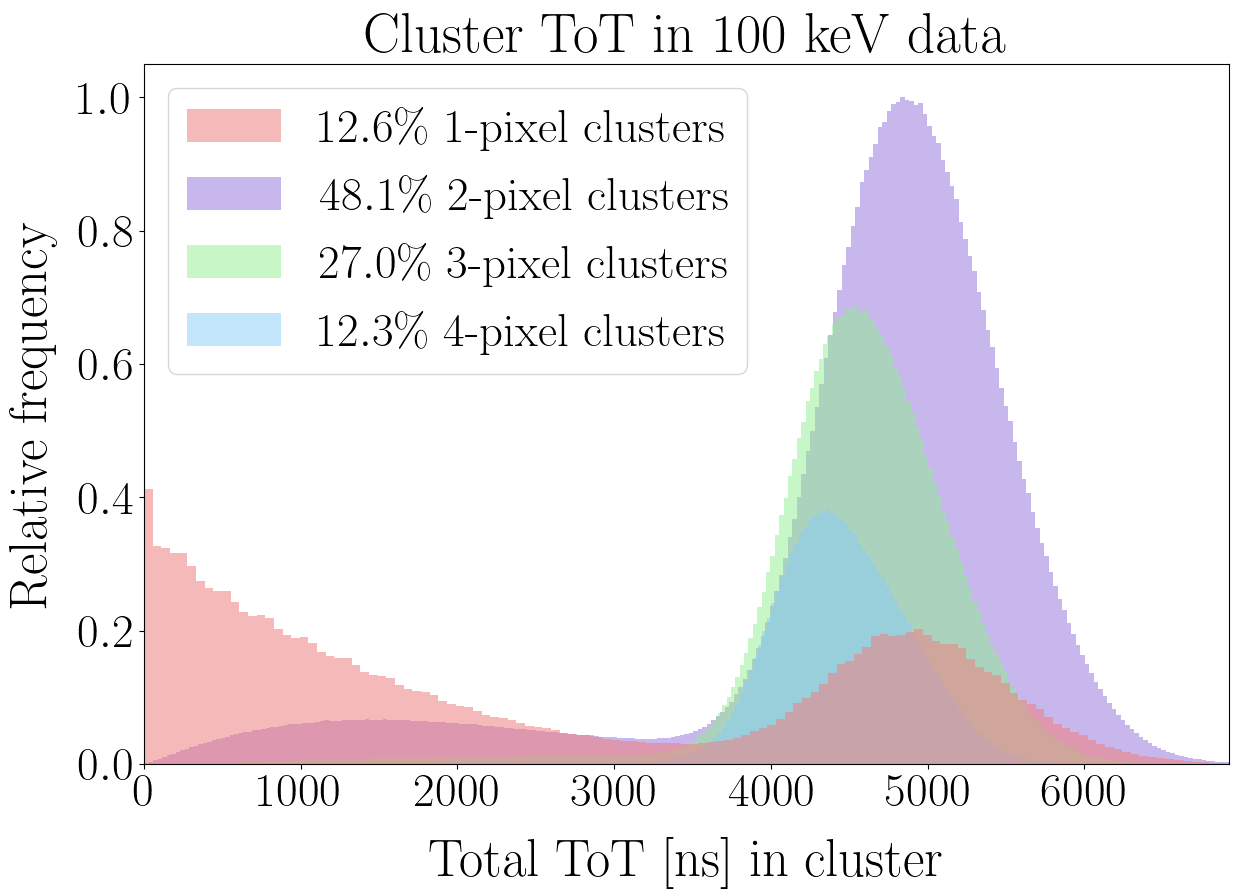}}
    \hfill
    \subfloat[]{\includegraphics[width=0.5\textwidth]{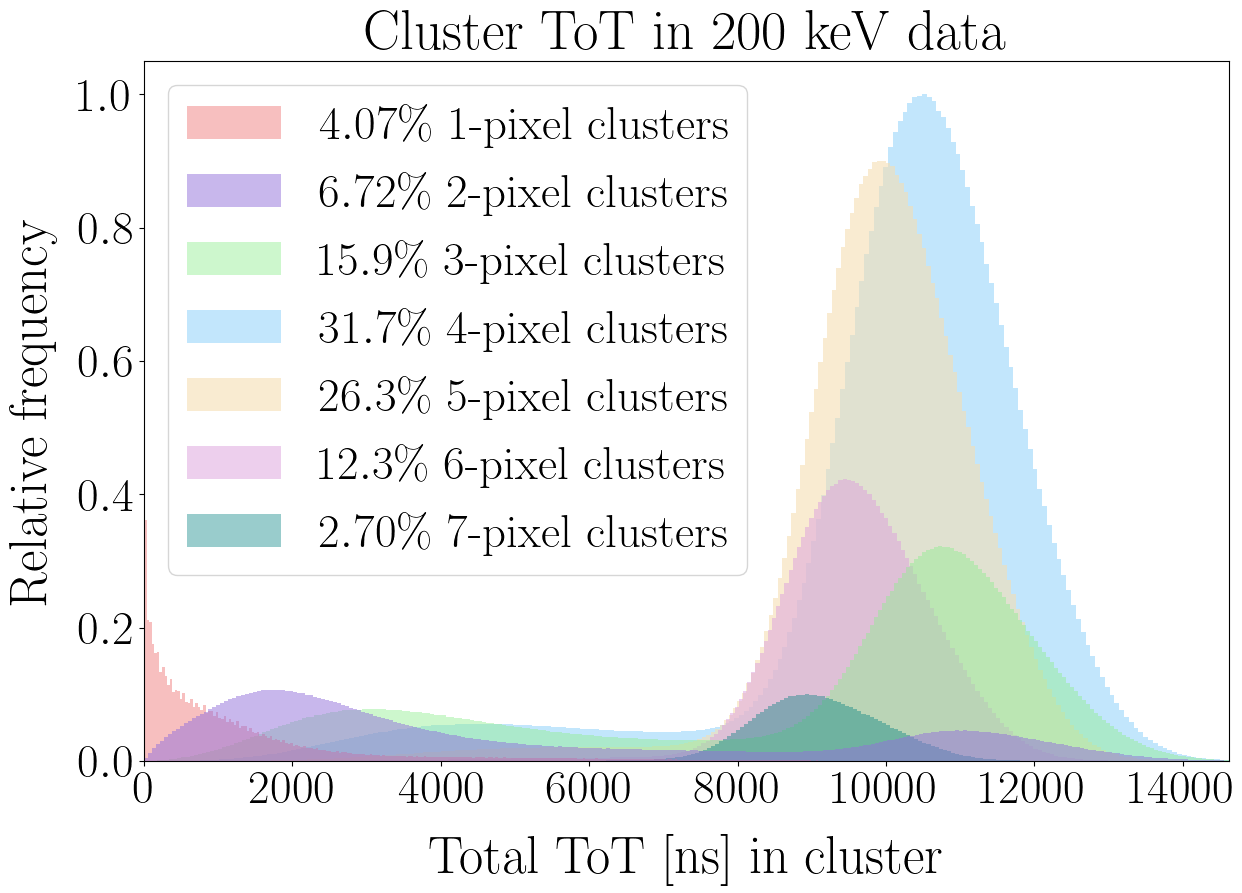}}
    \vfill
    \subfloat[]{\includegraphics[width=0.8\textwidth]{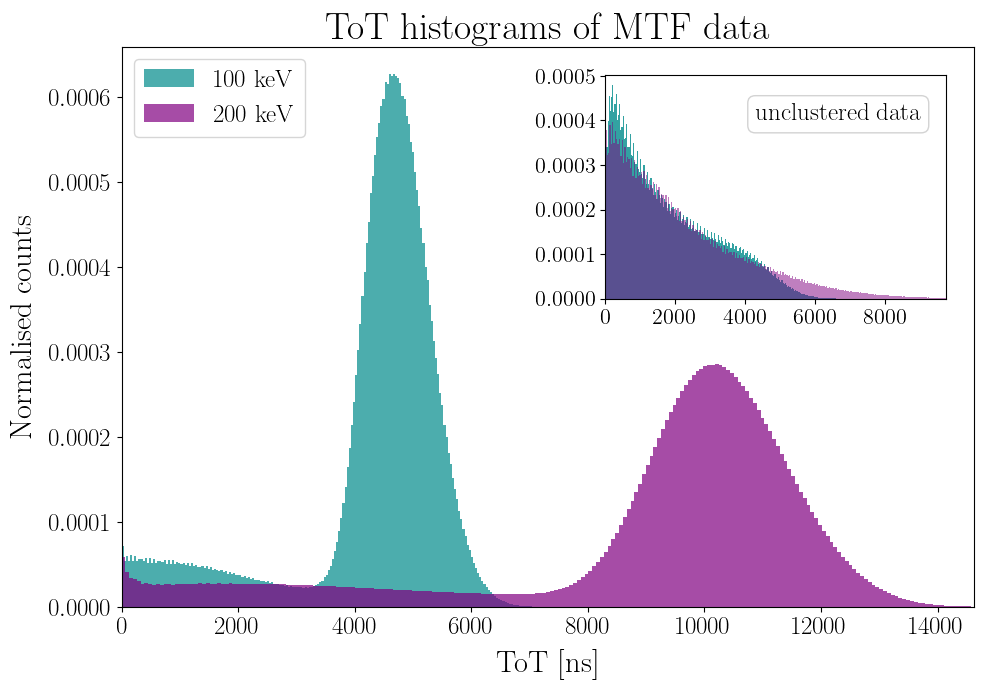}}
	\caption{Histograms of the total ToT per cluster in the data used for MTF calculation. Typical clusters due to (a) $100~$keV electrons consist of one to four pixels, whereas clusters due to (b) $200~$keV electrons consist of two to six pixels. The positions of the peaks for different cluster sizes depend on the energy lost due to charge sharing. Histograms of the ToT in all clusters (c) show the difference between the unclustered and clustered data.}
	\label{F10}
\end{figure}

An analysis of the spatial cluster size, which represents the number of pixels registering an event above threshold for a single electron impact, for both $100~$keV and $200~$keV electrons is presented in Figures~\ref{F10}a and ~\ref{F10}b. These clusters typically have spatial footprints ranging between one and four pixels for $100~$keV electrons, and between two and six pixels for $200~$keV electrons (refer to the interaction volumes in Figure~\ref{F2}). Additionally, larger footprints, ranging from five to fifteen pixels, have also been observed.

The Timepix4 data also includes the ToT information on charge deposited in a pixel. The distribution of this is shown in Figure~\ref{F10} as characteristic energy spectra for the 100 keV and 200 keV electrons.  Two phenomena due to the relatively high threshold set during data-taking, and the intrinsic spatial sampling of pixel detectors, are observed in these plots.

First, the ToT distributions for smaller clusters (clusters involving fewer hit pixels) exhibit a `tail' towards lower energies. This is a result of the entry point of an electron being close to the edge of a pixel. The electron's energy can be divided almost evenly between two pixels one of which passes threshold, and the other does not. This results in a lower fraction of the electron's energy being recorded, as a single-pixel cluster. The same applies to clusters of all sizes, however, for larger clusters, this `lost' energy comprises a smaller proportion of the total energy and so the ToT distribution remains closer to the expected for fully-contained events Gaussian. Additionally, it should be noted that this `lost' energy is not constant, but only bounded by the threshold value (approximately $10~$keV) resulting in broad, non-Gaussian distributions.

Secondly, there is a shift in the Gaussian peak position that is observable with increasing cluster sizes. The main contribution to this shift is the size of the threshold -- a cluster of size $n$ will be missing $n~\times~$the threshold energy. However, this is not the only contribution -- the process described above affects different-size clusters differently, so it is not possible to localize the ToT peaks with precision, without performing a per-pixel energy calibration.

Overall ToT spectra for the raw and clustered MTF data are presented in Figure~\ref{F10}c. As a result of the discussion above, the peaks are wide, with some low-energy components, and the positions of their means are only indicative of the overall detector performance, but here are not used quantitatively.  It would be possible to explore this further with a per-pixel energy calibration and missing energy cluster reconstruction correction that would result in narrower peaks, but this is not required for the MTF measurement presented here.

In order to improve the imaging resolution of the detector, clusters were replaced by centroids. This is a standard practice both in X-ray imaging~\cite{X_CLU}, where a cluster larger than a single pixel arise due to charge sharing, and in heavy particle tracking~\cite{E_CLU}, where large clusters are created by the particle traversing the bulk of the sensor. In the case of X-rays, ToT-based centroiding can reliably reconstruct an entry point for the photon. This is not always true for particle tracks. A charged particle generates ionization charge along its entire trajectory, in some cases depositing most of its energy at the end of its track in the silicon. This can cause a ToT-based centroid to be displaced from the actual entry point at the beginning of the track. Despite this, using a ToT-based centroid provides a more accurate estimate of the entry point than considering the entire cluster, and allows us to assign the centroid to a virtual pixel, smaller than the physical one, thereby still improving resolution.

\begin{figure}[!h]
	\centering
	\includegraphics[width=\textwidth]{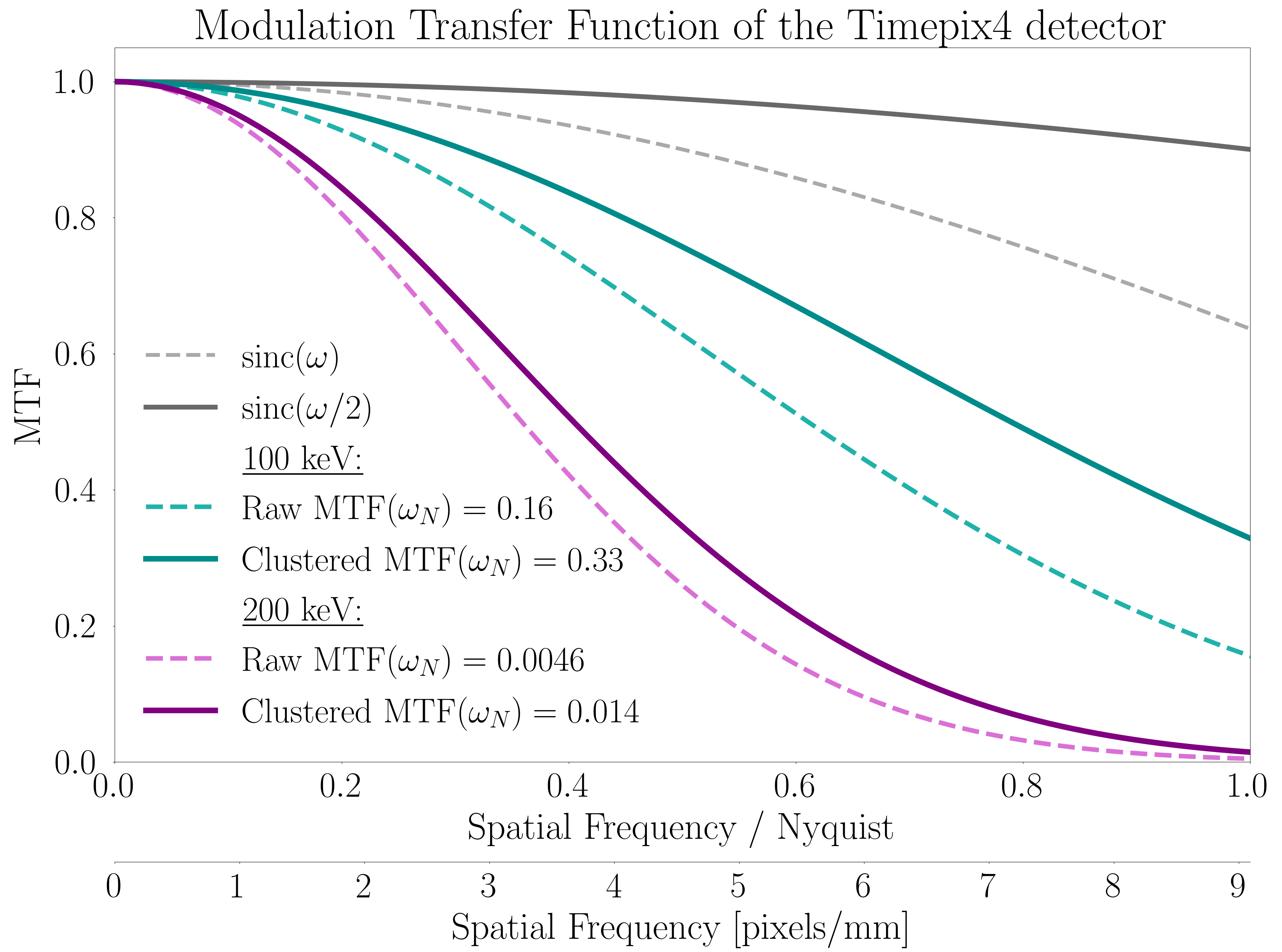}
	\caption{Electron MTF curves before and after clustering. The data was taken with $100~$keV and $200~$keV electrons for a slant-edge angle of $\theta=4^{\circ}$. The sampling region included $350$ pixels. The spatial frequency is in units of Nyquist frequency of the physical detector. The MTF curve for a perfect pixel with half the Timepix4 pixel pitch is also shown in solid grey.}
	\label{F7}
\end{figure}

The MTF for the detector was recalculated using the data clustered into  quarter-pixel-size virtual pixels, revealing an improvement factor of $2.12$ at the Nyquist frequency for $100~$keV electrons and $3.16$ for $200~$keV electrons (Table~\ref{T1}). The MTF curves calculated are shown in Figure~\ref{F7} and demonstrate the improvement over the unclustered data.

\begin{table}[!h]
\centering
\caption{The value of the MTF at Nyquist improves substantially after clustering.}
\smallskip
\begin{tabular}{c | c c | c}
    \hline
    Data & Raw & Clustered & Improvement \\
    Energy, (angle) & MTF at Nyquist & MTF at Nyquist & factor \\
    \hline
    $100~$keV, ($4^{\circ}$) & $0.16 \pm 0.0061$ & $0.34 \pm 0.010$ & $2.12 \pm 0.020$ \\
    $200~$keV, ($4^{\circ}$) & $0.0046 \pm 0.000069$ & $0.014 \pm 0.00015$ & $3.16 \pm 0.018$ \\
    \hline
\end{tabular}
\label{T1}
\end{table}

\begin{figure}[!h]
	\centering
    \subfloat[]{\includegraphics[width=0.495\textwidth]{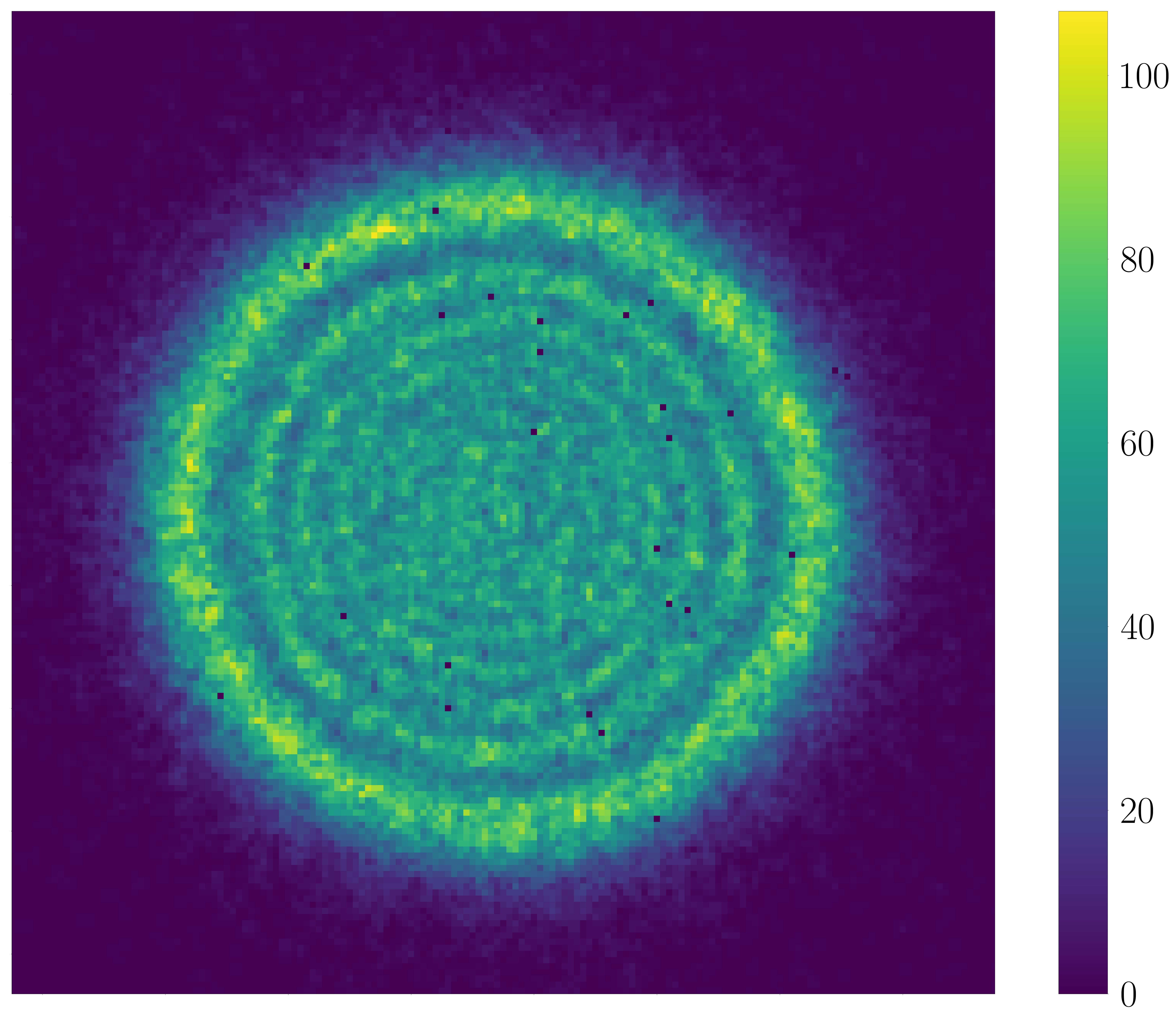}}
    \hspace{0.3em}
    \subfloat[]{\includegraphics[width=0.485\textwidth]{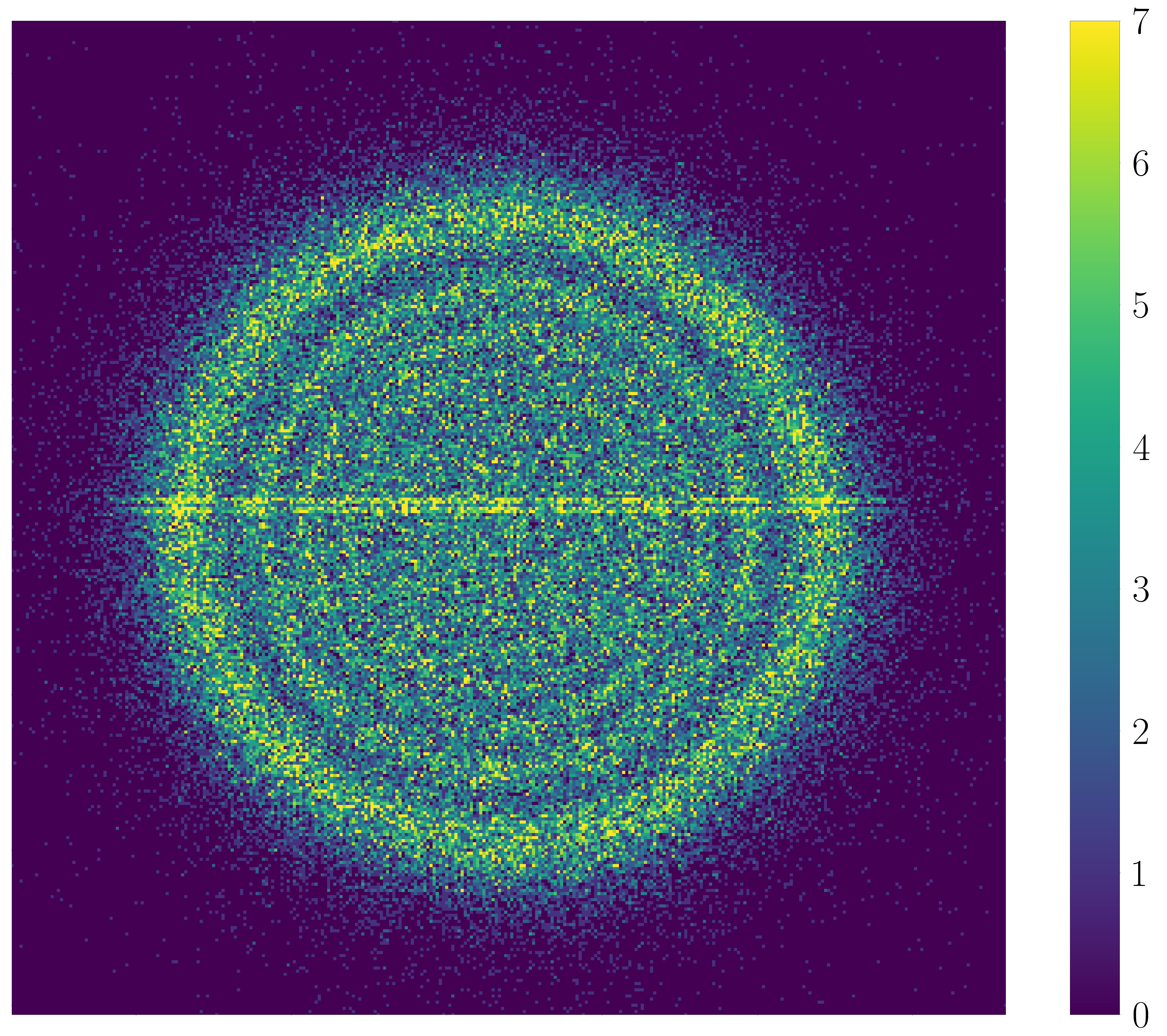}}
	\caption{Fresnel disk images which demonstrate the improvement in contrast achieved by the clustering procedure. (a) Raw images obtained with the Timepix4, and (b) the particle-clustered images, re-plotted into quarter-pixel-size virtual pixels. The line in the centre is due to a timing offset in the readout.}
	\label{F8}
\end{figure}

Figure~\ref{F8} shows test data acquired at the Rosalind Franklin Institute with $200~$keV electrons. The sub-pixel clustering results in a significant visible improvement in the image contrast and clarity. The clustered image exhibits an effect due to the two halves of the pixel matrix being read out in separate data streams. A misalignment between the timestamps of the two halves leads to a visible in the final image line along rows $255$ and $256$.

\clearpage

\section{Discussion}

This paper presents first results of the Timepix4 being exposed to electrons. There is considerable potential to further enhance the resolution achieved through clustering of the electron data. One specific improvement would be to implement a comprehensive calibration for the ToT-energy conversion and the ``timewalk" effect due to amplifier rise time. This would enhance the accuracy of the ToT-to-energy conversion and Time of Arrival (ToA) values, leading to a more precise input for the clustering process, and fully utilise the superior timing capabilities of the Timepix4 compared to previous generations.

Additionally, there is significant opportunity to optimize the electron clustering algorithm itself. This optimization will involve using sensor field simulations, Monte Carlo simulations of particle interactions within the sensor, and ASIC response simulations conducted with the Allpix2 package~\cite{A2}. These methods will help refine the algorithm's performance against known electron entry points, aiming to replicate and improve results achieved by implementing machine learning~\cite{vS}. The extensive MTF knife-edge dataset reported in this article will serve as a basis for testing these clustering optimizations.

Given the large datasets produced by a data-driven detector like the Timepix4, compared to particle counting detectors such as Medipix3, it is essential to implement online clustering during data collection. This could be accomplished using either an FPGA or a GPU, necessitating further adaptations to the clustering algorithms.

\section{Conclusions}

We conducted spatial resolution measurements of the Timepix4 hybrid pixel detector using the knife-edge MTF method. Data was taken in a TEM with electrons at $100~$keV and $200~$keV, and the MTF curves obtained from the unprocessed data were shown to be consistent with previous measurements on detectors with the same sensor type and pixel pitch~\cite{TEM1, TEM2, TEM3}. By applying a simple clustering algorithm that utilizes ToT-based centroiding, we achieved a notable improvement in the MTF at Nyquist frequency, enhancing it by factors of $2.12$ for $100~$keV and $3.16$ for $200~$keV. Test images clearly demonstrate this enhancement in contrast and clarity. Those results were obtained before conducting a large campaign of time characterisation or precise energy calibration. Continued development of this clustering technique is anticipated to further elevate the resolution of the Timepix4 for TEM applications.

\section*{Acknowledgments}

We would like to acknowledge significant contributions from: K. Paton for discussions on her work on the electron MTF measurements of Medipix3. The Medipix4 Collaboration for developing the Timepix4 detector and assisting with the development and operation. The Science and Technology Facilities Council for funding the CASE studentship. A. Knight, D. Yborde, M. Akhtar at Oxford's OPMD for wire bonding and engineering support.




\end{document}